\documentclass[review]{elsarticle}

\journal{Journal of \LaTeX\ Templates}

\usepackage{graphicx}
\usepackage[ruled, linesnumbered, vlined, lined, boxed]{algorithm2e}
\usepackage{todonotes}
\usepackage{arydshln}
\usepackage{url}
\usepackage{float}

\usepackage{footnote}
\usepackage{lineno}
\usepackage{multirow}
\usepackage{amsmath}

\usepackage[]{hyperref}
\usepackage[normalem]{ulem}

\newcommand{\SA}{\ensuremath{\mathsf{SA}}\xspace}
\newcommand{\A}{\ensuremath{\mathsf{A}}\xspace}
\newcommand{\B}{\ensuremath{\mathsf{DA}}\xspace}
\newcommand{\LF}{\ensuremath{\mathsf{LF}}\xspace}

\newcommand{\ISA}{\ensuremath{\mathsf{ISA}}\xspace}
\newcommand{\rank}{\ensuremath{\mathsf{rank}}\xspace}

\newcommand{\DA}{\ensuremath{\mathsf{DA}}\xspace}
\newcommand{\BWT}{\ensuremath{\mathsf{BWT}}\xspace}

\newcommand{\eg}{{\it e.g.}\xspace}

\newcommand{\C}[0]{\ensuremath{\mathsf{C}}\xspace}

\def\gSACAK{{\sf gSACA-K}\xspace}

\newcommand{\BIT}{\ensuremath{bit}\xspace}

\newcommand{\highlight}{\textcolor{black}}

\newcommand{\purple}{\textcolor{black}}

\newcommand{\bitvector}{\mbox{\textsc{bit\_plain}}\xspace}
\newcommand{\bitvectorsdv}{\mbox{\textsc{bit\_sd}}\xspace}
\newcommand{\ours}{\mbox{\textsc{Alg. 1}}\xspace}

\newcommand{\T}{\mathcal{T}}

\begin{document}

\begin{frontmatter}

\title{A simple algorithm for computing the document~array}


\author{Felipe A. Louza\corref{mycorrespondingauthors}}
\cortext[mycorrespondingauthors]{Corresponding author}
\ead{louza@ufu.br}

\address{Faculty of Electrical Engineering, Universidade Federal de Uberl\^andia,\\ Uberl\^andia, Brazil}

\begin{abstract}
We present a simple algorithm for computing the document array given a string
collection and its suffix array as input.
Our algorithm runs in linear time using 
\highlight{constant additional space for strings from constant alphabets.}
\end{abstract}

\begin{keyword}
Document array \sep Text indexing \sep Algorithms
\end{keyword}

\end{frontmatter}


\section{Introduction}

The {\em suffix array} (\SA)~\cite{Manber1993} is a fundamental data structure
in string processing that is commonly accompanied by the {\em document
array}~\cite{Muthukrishnan2002} when indexing string collections
(\eg~\cite{ValimakiM07,BelazzouguiNV13,KopelowitzKNS14,GagieHKKNPS17,SirenGNPD18}).
Given a collection of $d$ strings of total length $N$, the document array is an
array of integers $\DA[1,N]$ in the range $[1,d+1]$ that gives which document
each suffix in the suffix array belongs to.

It is well-known that \DA can be represented in a compact form by using 
a bitvector $\BIT[1,N]$ with support to rank operations,
requiring $N + o(N)$ bits of space~\cite{Sadakane07}.
However, there are applications where \DA must be accessed sequentially
(\eg~\cite{Ohlebusch2010,Arnold2011,Tustumi2016,LouzaTGZ19,EgidiLMT19,GuerriniR19}), and having the array $\DA[1,N]$ computed explicitly
is important.

In this paper we show how to compute \DA given a string collection and its
suffix array as input in $O(N)$ time.
\highlight{
Our algorithm reuses the space of \SA to store auxiliary arrays 
used to compute \DA. 
\SA is reconstructed in its original space during \DA computation.}
\highlight{
The workspace of our algorithm, that is, the extra space used in addition to the input and output, is 
\purple{$O(\sigma \lg N)$ bits}, where $\sigma$ is the alphabet size.
Therefore, for constant alphabets, the workspace of our algorithm is constant\footnote{We assume a computer word size of $\lg N$ bits.}.
}

\section{Background}

Let $T$ be a string of length $n$, over an alphabet $\Sigma$ of size $\sigma$,
such that $T[n]=\$$ is an end-marker symbol that does not occur elsewhere in
$T$ and precedes every symbol of $\Sigma$.  
$T[i,j]$ denotes the substring from $T[i]$ to $T[j]$ inclusive, for $1 \le i
\le j \le n$.
A suffix of $T$ is a substring $T[i,n]$.
We define $\rank_c(T,i)$ as the number of occurrences of symbol $c$ in
$T[1,i]$.
The string $T$ is stored in \purple{$n \lg \sigma$} bits of space.

The suffix array (\SA)~\cite{Manber1993} for $T$ is an array of integers
in the interval $[1,n]$ that provides the lexicographical order of all suffixes
of $T$.
The inverted permutation of \SA, denoted as \ISA, is defined as $\ISA[\SA[i]]=i$.
\SA can be computed in $O(n)$ time using $O(\sigma \lg n)$ bits
of workspace~\cite{Nong2013}.
{Since, \SA and \ISA are permutations of $[1,n]$, the arrays \SA and \ISA use \purple{$n \lg n$~bits} of space each one.}

The Burrows-Wheeler transform (\BWT)~\cite{Burrows1994} of $T$ 
is obtained by sorting all $n$ rotations 
of $T$ in a conceptual matrix $\mathcal{M}$, 
and taking the last column $L$ as the $\BWT$.  
It can also be defined 
through the relation 
\begin{equation}\label{e:bwt}
\BWT[i] = T[\SA[i]-1 \mod n].
\end{equation}
\purple{
The \BWT can be computed directly (without \SA) in $O(n)$ time using $O(n \lg \sigma)$ bits of workspace~\cite{Munro2017,SepulvedaNN19}, or alternatively in $O(n^2)$ time in-place~\cite{Crochemore2013}.
}

The Last-to-First (\LF) mapping states that the $k^{th}$
occurrence of a symbol $c$ in column $L$ of $\mathcal{M}$ and
the $k^{th}$ occurrence of $c$ in the first column $F$ 
correspond to the same symbol in $T$.
Let $\C[c]$ be the number of symbols $c'<c$ \purple{in $T[1,n]$}.
We define 
\begin{equation}\label{e:lf_rank}
\LF(i,c)=\C[c]+\rank_c(\BWT, i).
\end{equation}
We use shorthand $\LF(i)$ for $\LF(i, \BWT[i])$.
$\LF(i)$ may be
computed on-the-fly in $O(\lg \sigma)$ time
querying a wavelet tree~\cite{Grossi2003} for the rank queries on Equation~\ref{e:lf_rank}. 
The wavelet tree requires additional \purple{$(n~+~o(n)) \lg \sigma$} bits of space.

\LF-mapping allows us to navigate $T$ right-to-left, given
$T[k]=\BWT[i]$, then $T[k-1]=\BWT[\LF(i)]$.
$T$ can be reconstructed backwards from $\BWT$ 
starting with 
$T[n]=\BWT[1]=\$$, and repeatedly applying \LF for $n$ steps.

\subsection{String collections}\label{s:collection}

Let $\T = T_1, T_2, \dots, T_d$ be a collection of $d$ strings of lengths $n_1, n_2, \dots, n_d$.
The suffix array of $\T$ is the \SA built for the concatenation of all strings
$T^{cat}[1,N]=T_1 T_2 \dots T_d\#$, with total size $N=\sum_{i=1}^{d}(n_1)+1$ and a new end-marker $\#<\$$.
\SA can be computed in $O(N)$ time using $O(\sigma \lg N)$ bits
of workspace~\cite{Louza2017c},
such that
an end-marker $\$$ from string $T_i$ is smaller than
a $\$$ from string $T_j$ iff $i<j$,
which is equivalent to using $d$ different end-markers as separators, \highlight{without increasing the alphabet size}.

The \BWT may also be generalized for string collections.
\BWT of $\T$ is obtained from \SA and $T^{cat}$ through the Equation~\ref{e:bwt}.
However, \LF-mapping through Equation~\ref{e:lf_rank} does not work for symbols $\$$, since 
the $k^{th}$ symbol $\$$ in column $L$ does not (necessarily) corresponds
to the $k^{th}$ symbol $\$$ in column $F$, 
in this case $\LF(i,\$)$ is undefined~\cite{SirenGNPD18}.

$\LF(i)$ can be pre-computed in an array $\LF[1,n]$ through Equation~\ref{e:lf_isa}
such that \LF still works for $\$$-symbols.
Given $\SA$ and \ISA, we have
\begin{equation}\label{e:lf_isa}
\LF[i]=\ISA[(\SA[i]-1)\mod n]
\end{equation}
The array \LF uses \purple{$N \lg N$} bits of space.

\subsection{Document array}

The document array (\DA) is an array of integers in the interval $[1,d+1]$ that
tells us which document $j \in \T$ each suffix in the \SA
belongs to~\cite{Muthukrishnan2002}.
We define
$\DA[i]=j$ iff suffix $T^{cat}[\SA[i],N]$ came from string $T_j\in \T$. 
$\DA[1]=d+1$ for the last suffix $T^{cat}[N,N]=\#$.
$\DA[1,N]$ uses $N \lg (d+1)$ bits of space.

The array $\DA[1,N]$ can also be represented using a {\em wavelet
tree}~\cite{Grossi2003}, within the same $N \lg (d+1)$ bits with additional 
functionalities~\cite{ValimakiM07}.
\DA can still be compressed using grammars when the string collection is
repetitive~\cite{NavarroPV11}.

\subsection{Related work}\label{s:related}

Given $T^{cat}$ and \SA, the document array $\DA$ can be constructed in $O(N)$ time using $N \lg
N$ additional bits to store \ISA,
such that $\DA[\ISA[i]]=j$ for $i=\ell_{j-1},\dots,\ell_{j}$, with $\ell_0=1$ and $\ell_j=\sum_{k=1}^{j}n_k$, see~\cite[Alg. 5.29]{Ohlebusch2013}.

\DA can also be computed in the same fashion as the text $T^{cat}$ 
is reconstructed 
from its \BWT.
Given $T^{cat}$ and \SA, we can compute $\ISA[1,N]$ and then array $\LF[1,N]$ (Equation~\ref{e:lf_isa}). 
\DA is obtained in $O(N)$ time during the \BWT inversion using $N \lg N$ bits of workspace, see~\cite[Alg. 7.30]{Ohlebusch2013}.
\highlight{
In particular, in Section~\ref{s:ours} we show an alternative algorithm that
reuses the space of \SA to compute \LF without \ISA. 
Our algorithm uses $O(\sigma \lg N)$ bits of workspace and reconstructs \SA during \DA computation.
}

\paragraph{Lightweight alternative}
\DA can be computed using a compact data structure composed by 
a bitvector $\BIT[1,N]$ with rank support operation.
\BIT is built over $T^{cat}[1,N]$, such
that 
\begin{equation}\label{e:bit}
\BIT[i]=1\mbox{ iff }T^{cat}[i]=\$ \mbox{ and }\BIT[i]=0 \mbox{, otherwise.}
\end{equation}
$\DA[i]$ can be obtained using \BIT and \SA as follows~\cite[Alg. 7.29]{Ohlebusch2013}:
\begin{equation}\label{e:rank}
\DA[i]=\rank_1(\BIT,\SA[i])+1,
\end{equation}
$\BIT[1,N]$ can be pre-processed in $O(N)$ time so that rank queries are supported
in $O(1)$ time using additional $o(N)$ bits~\cite{Munro1996}.
This procedure computes \DA in $O(N)$ time using $N+o(N)$ bits of workspace.

\section{Computing \DA}\label{s:ours}

In this section we show how to compute \DA given \SA built for $T^{cat}$.
Our algorithm runs in {$O(N)$ time} using  $O(\sigma \lg N)$ bits of workspace,
which is constant when $\sigma = O(1)$.

At a glance, we reuse the space of \SA to compute the \LF-array, which is used to
traverse $T^{cat}[1,N]$ from right-to-left applying the \LF-mapping $N$ times.
We compute $\DA[1], \DA[\LF^{1}(1)], \DA[\LF^{2}(1)], \dots, \DA[\LF^{N-1}(1)]$.
Starting with $doc=d+1$, each $\DA[i]$ receives  $doc$,
and whenever $T^{cat}[\SA[i]-1]=\BWT[i]=\$$, that is, $\LF[i] \in [2, d+1]$, $doc$ is decremented by one.

Recall that \purple{$\LF(i, \$)$ is undefined then}
we cannot traverse backwards $T^{cat}[1,N]$ with the \LF-mapping given 
by Equation~\ref{e:lf_rank}.
Alternatively, given the \BWT of $T^{cat}$ and an auxiliary array $\C[1,\sigma]$ initialized 
with $\C[c]$ equal to the number of symbols $c'<c$ in $T^{cat}[1,N]$, we can
pre-compute correct \LF entries for every position with a corresponding \BWT
symbol $c \neq \$$. 
For $i=1,\dots,N$, 
$\LF[i]=\C[\BWT[i]]$, and $\C[\BWT[i]]$ is incremented by one.
The resulting (incorrect) \LF-positions, corresponding to $\BWT[i]=\$$, will be in the interval $[2,d+1]$. 
These values
will be computed correctly by Algorithm~\ref{a:ours} on-the-fly
during the right-to-left $T^{cat}[1,N]$ traversal.

\paragraph{Algorithm~\ref{a:ours}}
The algorithm starts with \SA stored in $\A[1,N]$.
\highlight{We use $N \lg N$ bits to store $\A[1,N]$, and $N \lg (d+1)$ bits to store $\B[1,N]$.}
First, we overwrite \SA with the \BWT in $\A[1,N]$ (Lines 1-3).
\highlight{Then, we overwrite the \BWT with the \LF-array computed as described above (Lines 4-6).
Recall that positions with $\A[i]\in [2,d+1]$ are not correct.} 
\highlight{In the sequel, $\DA[1,N]$ is computed while \SA is reconstructed in the space of $\A[1,N]$ as follows.}
Initially, $pos=1$ and $doc=d+1$ (Lines 7-8).
At each step $i=N,\dots, 1$ (Lines 9-18), the value in $\A[pos]$ (corresponding to $\LF(pos)$)
is stored in a temporary variable (Line 10) and replaced by $\SA[pos]=i$
(Line 11), then $\DA[pos]=doc$ 
(Line 12).  
\highlight{
Whenever $\LF[pos] = tmp \in [2,d+1]$, $\BWT(pos)$ is a $\$$-symbol 
and we have to compute correctly its \LF-mapping.}
\highlight{
In particular, 
when we reach the first $tmp \in [2,d+1]$, we reach the BWT position 
corresponding to the $d^{th}$ $\$$-symbol in $T^{cat}$ (the last one), 
because we traverse $T^{cat}[1,N]$ right-to-left, 
and its correct \LF-mapping is $tmp=d+1$.
The next iteration we reach $tmp \in [2,d+1]$ we are at the BWT position corresponding to the $(d-1)^{th}$ $\$$-symbol in $T^{cat}$, and
$tmp=d$, and so on.}
Therefore, whenever \highlight{$tmp \in [2,d+1]$} 
we update $tmp$ with the correct \LF-mapping value stored in $doc$ (Line 14), and 
$doc$ is decremented by one for the next iterations (Line 15). 
The next step will visit position $pos=tmp=\LF(pos)$ (Line 17).
At the end, $\DA[1,N]$ is completely computed and \SA is
reconstructed in the same space of $A[1,N]$.

\begin{algorithm}[t]
\renewcommand{\to}{\mbox{\bf\xspace to \xspace}}
\newcommand{\downto}{\mbox{\bf\xspace downto \xspace}}
\SetNlSty{textbf}{}{}
\SetAlgoLined
\SetCommentSty{mycommfont}

\For{$i \leftarrow 1 \to N$}{
	$\A[i] \leftarrow T^{cat}[\A[i]-1 \mod N]$ \tcp*{$\A = \BWT$}
}

\For{$i \leftarrow 1 \to N$}{
	$\A[i] \leftarrow \C[\A[i]]\mbox{++}$\tcp*{$\A = \LF$}
}

$pos \leftarrow 1$;

$doc \leftarrow d+1$;

\For{$i \leftarrow N \downto 1$}{

	$tmp \leftarrow \A[pos]$\tcp*{$tmp = \A[pos] = \LF(pos)$}

	$\A[pos] \leftarrow i$\tcp*{$\A[pos] = \SA[pos]$}

	$\B[pos] \leftarrow doc$

	\If(\tcp*[f]{$\BWT(pos) == \$$}){$tmp \leq d+1$}{

		$tmp \leftarrow doc$; 

		$doc \leftarrow doc-1$;
	}

	$pos \leftarrow tmp$\tcp*{$pos = tmp = \LF(pos)$}

}

\caption{Computing $\DA$ from $T^{cat}$, $\SA[1,N]$ and $\C[1,\sigma]$.}
\label{a:ours}
\end{algorithm}

\paragraph{Theoretical costs}
The number of steps is $N$ and only array $\C[1,\sigma]$ was needed in addition to the input and output. 
Therefore, the algorithm runs in {$O(N)$ time}, using
$O(\sigma \lg N)$ bits of workspace.

\paragraph{Discussion}
We remark that one can use a standard suffix sorting algorithm
(\eg~\cite{Nong2013}) to compute the suffix array for $T^{cat}$, such that
$\$$-symbols are not considered different symbols in $T^{cat}$ (see
Section~\ref{s:collection}), then $\LF(i,\$)$ is well-defined and
Algorithm~\ref{a:ours} can be applied with line 14 commented. 
Notice that, in this case, during suffix sorting unnecessary comparisons may be
performed, depending on the order of the strings in the collection, what may
deteriorate the practical performance of suffix sorting (see~\cite{Louza2017c} for
details).

\section{Experimental results}

We compared our algorithm with the lightweight alternative 
described in Section~\ref{s:related}. 
We evaluated two versions of this procedure, using compressed (\bitvectorsdv) and plain bitvectors (\bitvector).
We used C++ and SDSL library~\cite{Gog2014a} version 2.0. 
The algorithms receive as input 
\highlight{the concatenated string ($T^{cat}$) and its suffix array (\SA)}, 
which was computed using \gSACAK~\cite{Louza2017c}.
Our algorithm was implemented in ANSI C. 
The source codes are available at \url{https://github.com/felipelouza/document-array/}.

The experiments were conducted on a machine with Debian GNU/Linux 8 64 bits OS
(kernel 3.16.0-4) with processor Intel Xeon \mbox{E5-2630} v3 20M Cache
$2.40$-GHz, $386$ GB of RAM and a $13$ TB SATA disk.
We used real data collections described in Table~\ref{t:dataset}.

\begin{savenotes}
\begin{table}[t]
\centering
\caption{
Datasets.
We used $32$-bits integers to store $\SA[1,N]$ when $N<2^{31}$ (2GB), otherwise we used $64$-bits. 
The document array $\DA[1,N]$ is stored using $32$-bits integers, since $d$ is always smaller than $2^{31}$.
Each symbol of $T^{cat}$ uses 1 byte.
}
\label{t:dataset}
\begin{tabular}{|lrrrrr|}
\hline
\textbf{Dataset} & $\sigma$ & \textbf{$N/2^{30}$} & \textbf{$d$} &  \textbf{$N/d$} &  longest string \\
\hline
\texttt{revision}  & 203  & 0.39      & 20,433      & 20,527    & 2,000,452      \\
\texttt{influenza} & 15   & 0.56      & 394,217     & 1,516     & 2,867         \\ \hdashline
\texttt{reads}     & 4    & 2.87      & 32,621,862  & 94        & 101            \\
\texttt{pages}     & 205  & 3.74      & 1,000       & 4,019,585 & 362,724,758    \\
\texttt{wikipedia} & 208  & 8.32      & 3,903,703   & 2,288     & 224,488        \\
\texttt{proteins}  & 25   & 15.77     & 50,825,784  & 333       & 36,805        \\
\hline
\end{tabular}

\begin{description}

\item[\texttt{pages}:]
repetitive collection from a snapshot of Finnish-language Wikipedia.
Each document is composed by one page and its revisions\footnote{\url{http://jltsiren.kapsi.fi/data/fiwiki.bz2}}.

\item[\texttt{revision}:]
the same as \texttt{pages}, except that each revision is a separate
document.

\item[\texttt{influenza}:]
repetitive collection of the genomes of influenza
viruses\footnote{\url{ftp://ftp.ncbi.nih.gov/genomes/INFLUENZA/influenza.fna.gz}}.

\item[\texttt{wikipedia}:]
collection of pages from English-language of
Wikipedia\footnote{\url{http://algo2.iti.kit.edu/gog/projects/ALENEX15/collections/ENWIKIBIG/}}.

\item[\texttt{reads}:]
collection of DNA reads from Human Chromosome 14 (library
1)\footnote{\url{http://gage.cbcb.umd.edu/data/index.html}}.

\item[\texttt{proteins}:] 
collection of protein sequences from Uniprot/TrEMBL 2015\_09\footnote{\url{http://www.ebi.ac.uk/uniprot/download-center/}}.

\end{description}
\end{table}

\end{savenotes}

Table~\ref{t:results} shows the running time (in seconds) and workspace (in KB)
of each algorithm.
The workspace is the peak space used subtracted by the space used for the input, $T^{cat}[1,N]$ and
$\SA[1,N]$, and for the output, $\DA[1,N]$.

\paragraph{Results}
\bitvector was the fastest algorithm in all tests.
\bitvector was $2.19$ times faster than \bitvectorsdv, and
\highlight{$5.73$ times} faster than \ours, on the average.
\bitvectorsdv was still \highlight{$3$ times} faster than \ours, which
shows that \ours is not competitive \highlight{in practice}. 
On the other hand, \ours was the only algorithm that kept the workspace constant, namely  $1$ KB for inputs smaller than $2^{31}$ (2 GB) and $2$ KB otherwise, \highlight{which correspond to the space used by the auxiliary array $C[1, \sigma]$ used to compute \LF.} 
The workspace of \bitvector and \bitvectorsdv were much larger,
\bitvector spent $0.16 \times N$ bytes, whereas
\bitvectorsdv spent $0.003 \times N$ bytes, on the average.

\begin{table}[t]
\caption{
Running time and workspace.
}
\resizebox{\textwidth}{!} {
\begin{tabular}{|l|rrr|rrr|}
\hline
\multicolumn{1}{|c}{\multirow{2}{*}{\textbf{Dataset}}} & \multicolumn{3}{|c}{Time (seconds)} & \multicolumn{3}{|c|}{Workspace (KB)} \\
\multicolumn{1}{|c}{}                         & \multicolumn{1}{|c}{\ours} & \multicolumn{1}{c}{\bitvector} & \multicolumn{1}{c}{\bitvectorsdv} & \multicolumn{1}{|c}{\ours} & \multicolumn{1}{c}{\bitvector} & \multicolumn{1}{c|}{\bitvectorsdv} \\ \hline
\texttt{revision}  & 60.88	& \textbf{11.74}    & 20.37           & \textbf{1} & 64,002    & 44     \\
\texttt{influenza} & 109.13 	& \textbf{20.48}    & 41.24           & \textbf{1} & 91,168    & 704    \\ \hdashline
\texttt{reads}     & 931.35	& \textbf{150.40}   & 549.65          & \textbf{2} & 470,389   & 38,980 \\
\texttt{pages}     & 762.91	& 141.99            & \textbf{141.25} & \textbf{2} & 613,341   & {4}      \\
\texttt{wikipedia} & 2,947.59	& \textbf{450.64}   & 1,054.08        & \textbf{2}  & 1,363,147 & {7,096}  \\
\texttt{protein}   & 7,007.87	& \textbf{1,211.13} & 2,899.63        & \textbf{2}  & 2,583,532 & {69,423} \\   \hline
\end{tabular}
}
\label{t:results}
\end{table}

\paragraph{Competing interests}
The author declare that there is no competing interests.

\paragraph{Acknowledgments}
\highlight{
We thank the anonymous reviewers for comments that improved the manuscript.
We thank Giovanni Manzini, Travis Gagie and Nicola Prezza for helpful discussions.
}

\paragraph{Funding}
F.A.L. was partially supported by the grants $\#$2017/09105-0 and $\#$2018/21509-2 from the S\~ao Paulo Research Foundation (FAPESP).

\end{document}